# Can crowdsourcing rescue the social marketplace of ideas?

Taha Yasseri[1,2,3,4]*, Filippo Menczer[5,6]

[1] School of Sociology, University College Dublin, Dublin, Ireland.
[2] Geary Institute for Public Policy, University College Dublin, Dublin, Ireland.
[3] Oxford Internet Institute, University of Oxford, Oxford, UK.
[4] Alan Turing Institute for Data Science and AI, London, UK.
[5] Observatory on Social Media, Indiana University, Bloomington, USA.
[6] Luddy School of Informatics, Computing, and Engineering, Indiana University, Bloomington, USA.

\* corresponding author: taha.yasseri@ucd.ie

In the late 1980s, Tim Berners-Lee invented the World Wide Web to facilitate information sharing and collaboration among researchers. Rapid and widespread adoption by academic and commercial entities followed in the 1990s. After the burst of the dot-com bubble at the dawn of the new millennium, several social web platforms deeply transformed the information ecosystem within a few years. Wikis, blogs, social tagging and sharing sites, online social networks, and microblogs all significantly lowered the costs of producing and sharing content. This transition was hailed as a harbinger of more democratic participation in the information and knowledge economy — the realization of the "marketplace of ideas." No project was a better icon of such utopian dreams than Wikipedia: the revolutionary vision of an encyclopedia that can be read, and more importantly, edited by anyone, surprised media scholars and practitioners. The potential of the largest online collaborative project in history was soon realized, and the quality of Wikipedia articles reached a level comparable to professionally edited encyclopedias[10] despite its reported limitations and biases. Facebook and Twitter appeared a few years later. Ten years into the millennium, democratic revolutions were credited to these social media platforms and phenomena such as the Arab Spring were referred to as the Facebook revolution.

It did not take long for this sweet story to take a bitter twist. Hate speech, misinformation, polarization, manipulation, extremism, and data privacy scandals became commonplace online. In January 2021, the spread of disinformation and conspiracy theories on social media led to one of the darkest moments in the history of American democracy, when the U.S. Capitol was attacked on January 6. In the same month, Wikipedia celebrated its 20th birthday, enduring the test of time as a reputable source of information. How is it possible that one crowd-based social web technology stands as the most successful example of collaborative and healthy information sharing, while the other is blamed for epistemic chaos? What can this paradox teach us as we strive to fix the broken marketplace of ideas?

Both Facebook and Twitter have announced community-based review platforms to address misinformation. Can these crowdsourcing approaches work on social media? We provide an overview of the potential affordances of community-based approaches to content moderation based on past research on collaborative systems such as Wikipedia and preliminary analysis of data from Twitter. While our analysis generally supports a community-based approach to content moderation, it also warns against potential pitfalls, particularly when the implementation of the new infrastructure focuses on crowd-based "validation" rather than "collaboration."

To be sure, there are many important differences between Wikipedia and social media platforms, including design elements, user motivations and characteristics. The business model of social media, with its emphasis on engagement over quality, is also contrasted with the consensus-based model of Wikipedia. However, a review of past research points to the network effects of content generation and diffusion as a key to understanding how community-based moderation could rescue the social media marketplace of ideas, provided there is a serious intention by the commercial platforms to promote a healthier information environment.

**The collaborative content generation model**
There are several successful community-based information-sharing endeavors. Citizen science projects, where nonexpert volunteers collaboratively contribute to scientific discoveries through community interaction, show that such models can be robust.[11] Wikipedia is arguably the most successful community-based information-sharing initiative and for sure the most researched. Despite its generally acknowledged reliability, any experienced Wikipedia editor can tell stories of "edit wars" in which they have found themselves. Wikipedia edit wars have been studied in detail, showing that in most cases they lead to a consensus after enough time.[24] Agent-based simulations have helped reveal the secret of consensus-reaching mechanisms in Wikipedia. Articles, as shared products of editorial activities, facilitate indirect interactions between editors with opposing opinions, who would not engage in any positive interactions otherwise.[22] These indirect interactions provide both mechanisms and incentives to bring editors closer to one another and reach a consensus. Even though the simulated models are calibrated and validated against empirical data from Wikipedia, at an abstract theoretical level they suggest that any crowd-based system can reach a consensus if there is a collaborative element, regardless of the level of existing polarization. The same models were used to examine the effects of banning editors with extreme opinions. Counterintuitively, the exclusion of extremists hinders the consensus-reaching process.[16] Moreover, a moderate amount of controversy is correlated with higher article quality[17] and higher completeness and neutrality ratings. Wikipedia is not free of bias and abuse, such as inauthentic and coordinated accounts that manipulate content. Over time, the platform has dealt with these threats by evolving a complex set of community rules and norms along with a hierarchy of user and administrator roles, privileges, and tools (including benign bots[23]).

**The homophily-based model**
As opposed to Wikipedia, where articles are meant for a diverse community, the content created and shared on social media is targeted to one's social connections. Studies have reported that social networks on Twitter[8] and Facebook[19] tend to be structured around ideologically homogeneous echo chambers (Fig. 1a) and that users embedded in partisan echo chambers are exposed to more low-credibility information.[14] Homophily, the tendency to connect with like-minded individuals, leads to homogeneous communities both online and offline. However, social media mechanisms make it very easy to create as well as to cut off a social tie. Platforms encourage us to "block" users whose content we dislike. Mathematical analysis[4] and agent-based simulations[18] show that such costless unfollowing accelerates the formation of segregated echo chambers. Influence and social pressure ensure strong homogeneity of opinions within an echo chamber. One can generate engagement and gain popularity more easily by addressing a homogeneous audience than by appealing to common ground across diverse communities.

Another important ingredient of social media is algorithmic bias. Newsfeed ranking algorithms favor content that is likely to trigger engagement. Popularity also plays a role in Wikipedia, where content tends to be generated around topics that attract attention.[5] But on social media, engagement is directly amplified by the platforms as it is a crucial element in their business model. This leads to greater exposure to low-quality content[6] as well as ideologically aligned content that confirms or reinforces existing beliefs[13]. Algorithmic bias also increases vulnerability to manipulation by inauthentic or coordinated accounts. The incentives created by all of these social media characteristics — targeting homogeneous audiences as well as amplification of low-quality, confirmatory, and manipulated content — go against reaching a consensus outside of one's group. Worse, hate speech and verbal abuse are natural byproducts of this highly polarized information ecosystem and can spill over to real-world violence.[9]

**Community-Based Solutions**
Ronald Reagan once said: "Peace is not the absence of conflict, it is the ability to handle conflict by peaceful means." The current social media approach of hiding conflict through self-selection, unfollowing, removal of content, and account bans neither prevents nor mitigates conflict. Some researchers suggested exposing users to counter-attitudinal content, a positive algorithmic bias designed to break the bubbles. However, experiments show that partisan users become more entrenched in their beliefs once they are exposed to opposing views.[3] Moreover, mathematical models suggest that no matter how much positive algorithmic bias is applied, as long as unfollowing is costless, echo chambers will inevitably form.[4] What is effective, in contrast, is sharing personal experience[12] and —if we have learned one thing from the Wikipedia experience— collaborative interaction. Such collaboration in the context of social media could be aimed at tackling misinformation and community policy violations. Recent experiments suggest that crowdsourced layperson judgments can be effective at identifying misinformation.[15] Such a community approach could scale up fact-checking and moderation practices while mitigating both misinformation and polarization.[1]

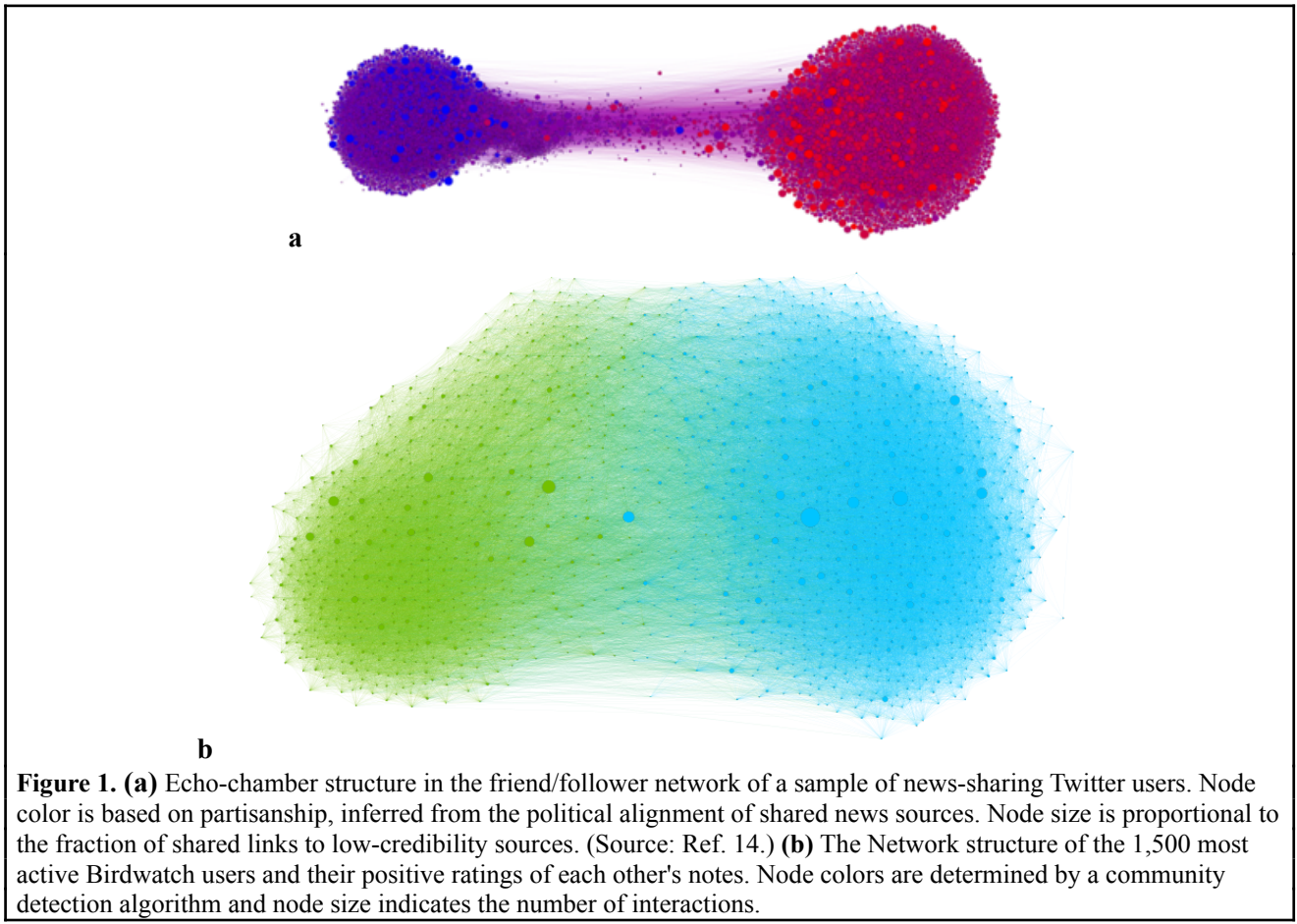

**Figure 1. (a)** Echo-chamber structure in the friend/follower network of a sample of news-sharing Twitter users. Node color is based on partisanship, inferred from the political alignment of shared news sources. Node size is proportional to the fraction of shared links to low-credibility sources. (Source: Ref. 14.) **(b)** The Network structure of the 1,500 most active Birdwatch users and their positive ratings of each other's notes. Node colors are determined by a community detection algorithm and node size indicates the number of interactions.

Following this line of argument, Facebook announced a community review program[21] in December 2019 and Twitter launched a community platform to address misinformation[7] in January 2021. Here we focus on Twitter's platform, originally called Birdwatch and later renamed *Community Notes* by Elon Musk, for which some preliminary data is available. In the current implementation, a member of the group of reviewers (selected by Twitter based on undisclosed criteria) can add a *note* to a tweet that they find "misinformed or potentially misleading." A note provides some information selected from predefined values about the tweet (misleading factual error, misleading satire) as well as some free text where the reviewer can comment and link to external sources. Then other reviewers express their agreement or disagreement with the existing notes through additional annotations such as helpfulness and informativeness. Ultimately, notes produced by reviewers will become visible next to the corresponding tweets based on the support/opposition they have received from other reviewers.

We analyzed Birdwatch helpfulness ratings as of December 2022 — 586,462 ratings of 31,422 notes by 15,023 reviewers. We observed evidence of a highly polarized network with two clusters where reviewers agree with those in the same group and disagree with those in the opposite group. To quantify polarization among reviewers, we considered pairs of reviewers who rated each other at least once. Of these pairs of reviewers with reciprocal ratings, 60-70% are consistent in that both rate each other as helpful (in the same cluster) or not helpful (in different clusters). Furthermore, we looked at triads of reviewers with reciprocal ratings to quantify structural balance, which occurs when all three reviewers agree with each other (in the same cluster) or when two reviewers agree with each other (in the same cluster) and disagree with the third (in the other cluster). Even though 39% of ratings are negative, we found that only 25% of triads are structurally imbalanced. This indicates a high level of cohesion within the clusters and disagreement across the clusters when interactions exist. Fig. 1b offers visual confirmation of these findings by mapping the network of Birdwatch reviewers. An edge between two nodes represents reciprocal ratings that indicate agreement on average. The polarization among Birdwatch reviewers mimics the one observed among generic Twitter users (Fig. 1a).

It is unlikely that this polarization is a reflection of objective arguments; rather, it merely represents the political affiliations of the reviewers. Analysis of the notes confirms that users systematically reject content from those with whom they disagree politically[2]. One might argue that the population of Birdwatch reviewers is less homogenous than that of Wikipedia editors. This may be true, yet a polarized crowd can be even more effective in producing high-quality content compared with a homogenous team.[20] The missing ingredient, however, is collaboration: reviewers of opposing opinions currently do not have to reach a consensus. The design of community review systems will have to be modified to enforce collaboration rather than competitive behavior; robustness to competition is as critical as resistance to coordinated manipulation. Wikipedia teaches us that community rules can enforce such norms.

**Conclusion**

Solutions to the epistemic chaos brought about by social media platforms will have to involve multiple dimensions and stakeholders, including digital media literacy and government regulation. Past research on Wikipedia and social media platforms has shown that collaborative content moderation and network effects are an important part of the equation. Collaboration, civilized challenge, and mediated controversy are keys to the success of Wikipedia, while platform-amplified engagement focused on homogeneous audiences contributes to clash and abuse — online and offline. Our preliminary analysis of data from Twitter's community-based review system suggests a design that does not afford collaboration among a diverse crowd. The situation at Twitter has been changing rapidly since Musk's takeover and it is hard to predict what the platform will look like when this article arrives in print. Regardless of this volatility or the specific futures of Twitter, the lessons discussed here are important for the social marketplace of ideas, whether on Twitter or distributed platforms such as Mastodon. In particular, while crowdsourcing initiatives may present viable options to rescue the marketplace, collaborative designs are crucial; direct collaboration seems to be a key element in successful crowd-based systems.

Technological advancements in web-supported collaboration are not yet fully exploited in the area of content moderation and a dire need for innovative designs is apparent. The unit of content analysis could be more than a single tweet or post and a more holistic, conversation-based consensus building could be the goal of the content moderators. The quality of conversations can be determined and enhanced by a diverse set of moderators and this may facilitate a quality-based promotion system instead of the current ones that are mainly based on engagement. In addition to providing a better-functioning content moderation system, these features will hinder astroturfing and other tools of social media-based propaganda.

**Acknowledgements**
We are grateful to David G. Rand, Janos Kertesz, Ksenia Musaelyan, and Dimitar Nikolov for their feedback and comments on the manuscript. TY is partially supported by Google. FM is supported in part by Knight Foundation and Craig Newmark Philanthropies.

**Competing Interest Statement:** The opinions and recommendations expressed here are those of the authors and do not necessarily reflect the views of the funders. The authors declare no competing interests.

# Authors

**Taha Yasseri** (taha.yasseri@ucd.ie) is an associate professor of Sociology and a Geary Fellow in Public Policy at University College Dublin, Dublin, Ireland.

**Filippo Menczer** (fil@indiana.edu) is the Luddy Distinguished Professor of Informatics and Computer Science and the director of the Observatory on Social Media at Indiana University, Bloomington, IN, USA.